\documentclass[journal=langd5,manuscript=article]{achemso}
\usepackage[version=3]{mhchem}

\author{Alexandre P. dos Santos}
\email{alexandre.pereira@ufrgs.br}
\affiliation{Departamento de F\'isica, Universidade Federal de Santa Catarina, 88040-900, Florian\'opolis, Santa Catarina, Brazil}

\author{Wagner Figueiredo}
\email{wagfig@fisica.ufsc.br}
\affiliation{Departamento de F\'isica, Universidade Federal de Santa Catarina, 88040-900, Florian\'opolis, Santa Catarina, Brazil}

\author{Yan Levin}
\email{levin@if.ufrgs.br}
\affiliation{Instituto de F\'isica, Universidade Federal do Rio Grande do Sul, Caixa Postal 15051, CEP 91501-970, Porto Alegre, RS, Brazil}

\title{Ion Specificity and Micellization of Ionic Surfactants: A Monte Carlo Study}

\begin{document}

\begin{abstract}

We develop a simulation method which allows us to calculate the critical micelle concentrations for ionic surfactants in the presence of different salts. The results are in good agreement with the experimental data. The simulations are performed on a simple cubic lattice. The anionic interactions with the alkyl chains are taken into account based on the previously developed theory of the interfacial tensions of 
hydrophobic interfaces: the kosmotropic anions do not interact with the hydrocarbon tails of ionic surfactants, while chaotropic anions interact with the alkyl chains through a dispersion potential proportional to the anionic polarizability.

\end{abstract}

\maketitle

\section{Introduction}

Micelles are important for various applications such as drug carriers for treatment of tumors~\cite{TrKo10,WeLi13}, as detergents and paints in the chemical industry, and emulsifiers in food industry. The process of micellization is, therefore, a subject of intense investigation~\cite{GuJo80,Ba01,GiHe06,HeFi09,JuLe12}. Similar to many other physicochemical systems, micellar formation is strongly influenced by the ions present in the solution~\cite{JiLi04,Ro07,LuMa11,MuMo12,MuDe13,MaKa13}. Hofmeister~\cite{Ho88} was the first to observe that different salts have a profoundly distinct effect on protein solutions~\cite{ReZh98,MeBa12,RePa12,ReGu13}, with anions affecting the stability of proteins more strongly than cations. Hofmeister's work resulted in the celebrated lyotropic (Hofmeister) series, a list of ions ordered by their ability to precipitate proteins. The series has been observed in many different fields of study such as biophysics~\cite{Co97}, colloidal science~\cite{LoJo03,LoSa08,PeOr10,CaFa11,ScNe12}, bacterial growth~\cite{LoNi05}, micelle-vesicle transitions~\cite{ReVl07}, ionic liquids~\cite{AoKi13,WaSu13}, surface tensions~\cite{AvSa76,CoWa85,WePu96,MaTs01,JuTo06,PeRe07,BaMu11,ToSt13,LiDe13}, peptide bonds~\cite{HeVi10}, liquid crystals~\cite{DaLa09,CaMa12}, microemulsions~\cite{MuMo04}, porous interfaces~\cite{CeMa13}, etc. Recently, a theory was developed which allowed to accurately calculate surface and interfacial tensions of electrolyte-air~\cite{LeDo09,DoDi10,DoLe10,DoLe13} and electrolyte-oil~\cite{DoLe12} interfaces. The theory showed that, near a hydrophobic surface, kosmotropes remain strongly hydrated and are repelled from the interface.  On the other hand, chaotropes can adsorb to the interface by polarizing their electronic cloud and thus gaining hydrophobic cavitational energy~\cite{Le09}. For the water-oil interface, the theory also showed the fundamental importance of dispersion interactions~\cite{MuMo04} between the ions and the hydrocarbons~\cite{DoLe12}. The same 
mechanism was also 
found to be responsible for the adsorption of hydrophilic cationic polyions to a hydrophobic 
wall~\cite{DoLe13b} and for variation of  the critical coagulation concentrations (CCCs) of hydrophobic colloidal suspensions in the presence of chaotropic anions~\cite{DoLe11,PeOr10}. 
These earlier calculations suggest that the interactions of chaotropic anions with the hydrocarbon tails of ionic surfactants will be predominantly controlled by the dispersion forces and should be proportional to the ionic polarizability.  On the other hand, the kosmotropic anions should remain strongly hydrated and
should not feel the dispersion interaction with the surfactant tails.

In the present paper we will explore the process of micellization of ionic surfactants in the presence of various Hofmeister electrolytes.  Our goal is to develop a simple model which will allow us to
calculate the critical micelle concentrations (CMCs) for solutions of ionic surfactants in the presence of different salts.
In the next section, we will briefly review a standard lattice model used to study CMCs of neutral 
surfactant molecules and discuss how this simple model can be modified to account for the surfactant 
charged head groups and for salt present in solution.

\section{Model and Monte Carlo Simulations}

The Monte Carlo simulations are performed on a three dimensional square lattice~\cite{GiFi00}, in which each cell represents a charged monomer, a neutral monomer, a water molecule, or an ion. An amphiphilic molecule is modeled as a charged head and three adjacent neutral monomers, representing the tail of the surfactant. The other species are monovalent counterions, cations, and anions. The cells are distributed in a cubic box of side $L$, defined by the concentration of surfactants, $\rho_t$. The typical number of surfactants is around $40$, while $L$ should be a multiple of the unit cell length, $4$~\AA, a typical ionic diameter. We consider periodic boundary conditions and the electrostatic interactions are calculated using the Ewald summation method~\cite{AlTi87}. In order to model the electrostatic interactions, we consider the water as an uniform dielectric with relative permittivity, $\epsilon_w = 80$. The Bjerrum length is defined as $\lambda_B=\beta q^2/\epsilon_w$, where $q$ is the proton 
charge, $\beta=1/k_BT$, $k_B$ is the Boltzmann constant, and $T$ is the absolute temperature. For water at room temperature, the Bjerrum length is $7.2$~\AA. In order to take into account the hydrophobic interactions, which result in  micellar formation, we follow the method developed for polar head surfactants~\cite{GiFi00}. The interaction energy between two adjacent tail monomers, from different surfactants, is taken to be $-\epsilon$, while between two adjacent tail monomers and an empty cell~(water) is $+\epsilon$. In order to avoid condensed surfactants~\cite{GiFi00}, we also consider that each bend in a surfactant molecule costs $+\epsilon$, and if the head and the last tail monomer are on the adjacent sites, it gives an additional energy penalty $+\epsilon$. 
This is a minimal model of micellar formation, with only one adjustable parameter $+\epsilon$.  The model can be
elaborated further  at a price of introducing more adjustable parameters.  Here, however, we are interested to 
explore if the minimal model is sufficient to quantitatively account for the CMCs of ionic surfactants in solutions containing various electrolytes.

\begin{figure}[t]
\begin{center}
\includegraphics[width=7.cm]{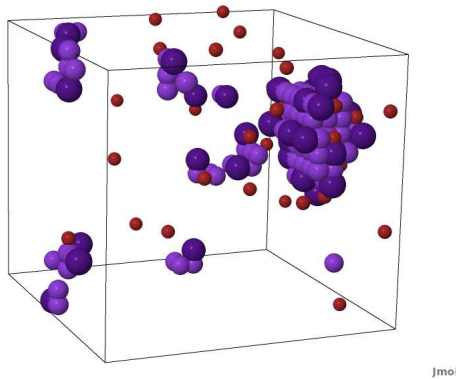}
\end{center}
\caption{A snapshot of a typical salt free equilibrium configuration. 
The surfactant concentration is $\rho_t=100$~mM. The small red spheres represent counterions, the larger purple spheres represent the surfactant tail monomers, and the largest spheres represent the charged head groups.}
\label{fig1}
\end{figure}

Starting from a random 
distribution 
of molecules, we apply the Metropolis algorithm for $1\times10^5$ Monte Carlo steps, until the system reaches  equilibrium. The movements of ions are simple one cell translations to a nearest neighbor site, while the 
surfactants can move by either a reptation or a
translation to a completely random place on the lattice, with $50-50\%$ probability for either move. After equilibration, we save $1\times10^5$ uncorrelated states. An uncorrelated state is achieved each $1\times10^4$ 
Monte Carlo steps. The data is then analyzed to obtain the concentration of ``free" surfactants $\rho_f$  as a function of the total surfactant concentration $\rho_t$. The free surfactants are defined as molecules which do not have tail monomers of other amphiphiles adjacent to theirs~\cite{GiFi00}. In Fig.~\ref{fig1}, we show a snapshot of a typical equilibrium configuration.  

\section{Results}

\begin{figure}[t]
\begin{center}
\includegraphics[width=7.cm]{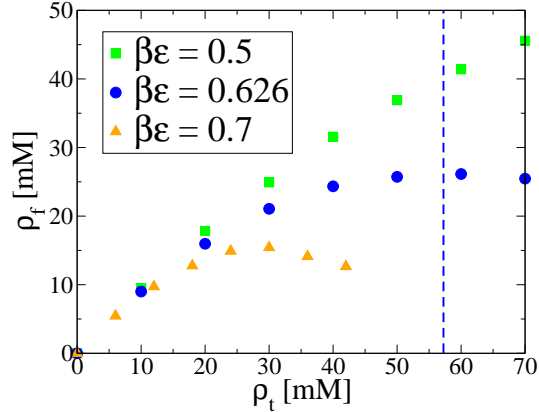}
\end{center}
\caption{Concentration of free surfactants as a function of the total surfactant concentration. The vertical dashed line represents the experimental~\cite{LuMa11} CMC, $57.2$~mM. Observe that it passes very near the maximum of the simulation curve with $\beta\epsilon=0.626$, at $59.2$~mM.  This value of $\beta\epsilon$ will then be used to model DMIM.}
\label{fig2}
\end{figure}

We first consider a salt free solution containing cationic surfactants and negatively charged counterions. A recent experiment~\cite{LuMa11} showed that for a cationic surfactant  1-decyl-3-methyl-1H-imidazolium chloride (DMIM), with no added salt, the CMC is around $57.2$~mM. Our first goal is to adjust $\epsilon$ to obtain the correct experimental CMC.  Prior to micellar formation, the concentration of free surfactants will increase with  the total concentration.  However, after the CMC, the new surfactants added to solution will go into micelles, and the concentration of free molecules will  decrease.  We will, therefore, define the CMC as the concentration $\rho_t^*$  at which $\rho_f$  is at maximum. In Fig~\ref{fig2}, we show that for 
$\epsilon=0.626~k_BT$, the maximum of the curve is at $\rho_t^*=59.2$~mM, very close to the experimental 
CMC. For $\epsilon=0.7~k_BT$, the maximum is around $29$~mM, while for $\epsilon=0.5~k_BT$, we cannot find a maximum in the range of the data, see Fig.~\ref{fig2}. The precise maximums of all the curves are obtained by interpolating the simulation data, with the cubic spline method.
We should note that the CMCs of salt free surfactant solutions depend strongly on the chain length~\cite{JuLu08}. In particular
the alkyl chains of most surfactants have more than 4 monomers used in our lattice model.  Nevertheless, it has been  
shown previously that the properties of formation of micelles are captured quite well with a lattice model 
containing only $4$ monomers and a suitably adjusted parameter~\cite{BrCa92} $\epsilon$. We will suppose that the same 
remains true in the presence of salt.

\begin{figure}[t]
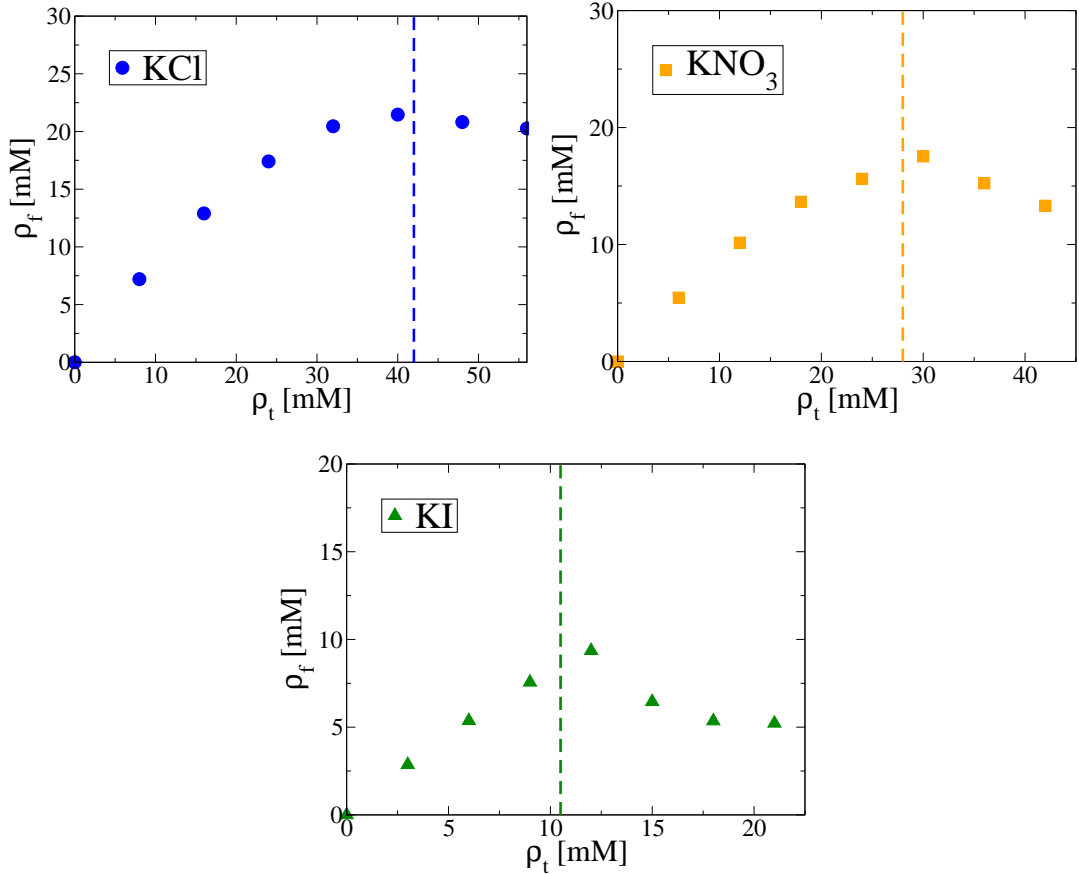

\begin{center}
\includegraphics[width=7.cm]{fig3a.eps}\vspace{0.2cm}\hspace{0.1cm}
\includegraphics[width=7.cm]{fig3b.eps}\vspace{0.2cm}\hspace{0.1cm}
\includegraphics[width=7.cm]{fig3c.eps}\vspace{0.2cm}\hspace{0.1cm}
\end{center}
\caption{Concentration of free surfactants as a function of $\rho_t$, with added $50$~mM salts. The dashed lines represent the experimental~\cite{LuMa11} CMC values for \ce{KCl}, \ce{KNO3} and \ce{KI}, $42$, $28$ and $10.5$~mM, respectively, in good agreement with the values obtained in simulations, $39.94$~mM, $29.93$~mM and $11.71$~mM, respectively.}
\label{fig3}
\end{figure}
The experiments also provide us with the values of CMCs for solutions containing various electrolytes~\cite{LuMa11}, \ce{KCl}, \ce{KBr}, \ce{KNO3} and \ce{KI}. Using the same $\epsilon=0.626~k_BT$, adjusted for the case without salt, we now calculate the CMCs for solutions containing electrolyte at concentration of $50$~mM. Consider first the case with added \ce{KCl}. 
Following the theory developed in~\cite{LeDo09,DoDi10,DoLe10,DoLe11,DoLe12,DoLe13} the small alkali metal cations and light halide anions, such as \ce{Cl-} and \ce{F-}, remain hydrated and do not interact with the hydrocarbons. The only interaction between the ions and surfactants is the steric (hardcore) repulsion with the monomers and Coulomb interaction with the head groups.
The CMC obtained from simulations for solution with $50$~mM \ce{KCl} is  $39.94$~mM, in a good agreement with the experimental value, $42$~mM, see Fig.~\ref{fig3}. 
\ce{KCl}  screens the electrostatic repulsion between the head groups, thus lowering the CMC.

We next consider salts with chaotropic anions, \ce{KNO3}, \ce{KI} and \ce{KBr}. The earlier work showed that chaotropic anions, such as \ce{NO3-}, \ce{I-} and \ce{Br-}, adsorbs to hydrophobic surfaces mainly due to dispersion interaction between the ions and the hydrocarbons. The dispersion potential is proportional to the ionic polarizability~\cite{DoLe11,DoLe12}.  As the simplest approximation we will, therefore, take the interaction potential between a chaotropic anion and a surfactant monomer to be a simple square well of one lattice spacing (nearest-neighbor interaction) with the depth proportional to the ionic polarizability $\gamma$,
\begin{equation}
\beta U_{c}=-\nu \gamma \ ,
\end{equation}
where  $\nu$ is an adjustable, ion independent, parameter. 

\begin{figure}[t]
\begin{center}
\includegraphics[width=7.cm]{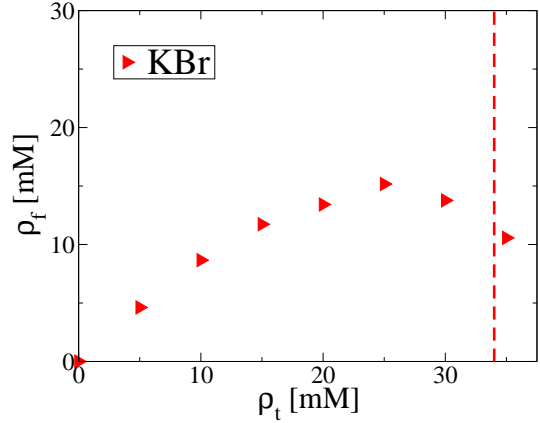}
\end{center}
\caption{Concentration of free surfactants as a function of the total surfactants concentration, for $50$~mM \ce{KBr}. The dashed line represents the experimental~\cite{LuMa11} CMC value for \ce{KBr}, $34$~mM. The CMC found in simulation is $25.48$~mM.}
\label{fig4}
\end{figure}

Consider first the salt $\ce{KI}$ (at $50$~mM). We will try to adjust the value of the parameter $\nu$ to obtain the correct experimental CMC, $10.5$~mM. The same value of $\nu$ will then be used for all other chaotropic anions. The ionic polarizabilities for \ce{I-}, \ce{Br-} and \ce{NO3-}, provided in Ref.~\cite{PyPi92}, are $\gamma=7.4$, $5.07$, and $4.09~$\AA$^3$, respectively, and are the same as 
used in the previous theoretical works~\cite{LeDo09,DoDi10,DoLe10,DoLe11,DoLe12,DoLe13} on surface and interfacial tensions. It is difficult to find the exact value of $\nu$ for $\ce{KI}$, however with $\nu=0.155~$\AA$^{-3}$ we obtain a CMC of $11.71$~mM, in a reasonable agreement with the experimental data, see Fig.~\ref{fig3}. Using the same value of  $\nu$, we can now calculate the CMCs for other salts, \ce{KNO3} and \ce{KBr}. For the salt \ce{KNO3} the value found from simulations is $29.93$~mM, and is very close to the experimental CMC of $28$~mM, see Fig.~\ref{fig3}. For the salt \ce{KBr} we obtain the value $25.48$~mM, while the experimental value is $34$~mM, see Fig.~\ref{fig4}.
The difference in the value of the CMC for \ce{NaBr} obtained in simulations and experiments is likely due to the overestimate of the ionic polarizability of \ce{Br-} anion~\cite{PyPi92}. This problem was already noticed in the  earlier work on the stability of hydrophobic colloidal suspensions~\cite{DoLe11}, and in the recent \textit{ab initio} simulations~\cite{BaLu10}. In Table~\ref{tab1} we summarize the CMCs of DMIM in the presence of various electrolytes at $50$~mM concentration.
\begin{table}[h]
\begin{tabular}{c|c|c}
      \hline
      \hline
                        & simulations~(mM) & experiments~(mM)~\cite{LuMa11} \\
      \hline                        
      no salt           &  59.2            & 57.2             \\
      \hline      
      Salts (at $50$~mM)  &                  &                  \\
      \ce{KCl}          &  39.94           & 42               \\
      \ce{KBr}          &  25.48           & 34               \\
      \ce{KNO3}         &  29.93           & 28               \\
      \ce{KI}           &  11.71           & 10.5             \\
      \hline
      \hline      
\end{tabular}
\caption{Summary of the calculated and experimental CMCs for DMIM surfactant. The results are for systems without added salt and with salts at $50$~mM. The experimental data are from Ref.~\cite{LuMa11}. The $\beta\epsilon$ and $\nu$ constants are adjusted to $0.626$ and $0.155~$\AA$^{-3}$, respectively.}
\label{tab1}
\end{table}

We next note that the Hamiltonian of the  lattice model with monovalent kosmotropic ions is invariant under the charge inversion $+ \rightleftharpoons -$  of all the ionic species.  This means that the CMCs of anionic surfactants with the same 
chain length, and a head group of roughly the same size, should be similar to the CMCs of cationic surfactants. 
Unfortunately, no experimental data for CMCs of anionic surfactants, with the same chain length as DMIM and with added salt, are available. 
However, for anionic surfactant sodium 9-decenyl sulfate~(SDeS) (with the same chain length as DMIM) the 
salt-free CMC measured experimentally is $61$~mM~\cite{AkSh04} --- very close to the CMC of  DMIM~\cite{LuMa11}, $57.2$~mM, showing that the charge inversion symmetry is indeed closely respected --- once again validating our model. 
For chaotropic anions preferential adsorption to alkyl chains results in an
increased  net charge of anionic surfactants, while for cationic surfactants anionic adsorption lowers the net charge of the surfactant molecule.
This breaks the charge inversion symmetry, resulting in different CMCs of cationic and anionic surfactants 
(of the same chain length) is solutions containing chaotropic ions. 
In Table~\ref{tab2}, we summarize the predictions of our model for 
CMCs of anionic surfactant sodium 9-decenyl sulfate~(SDeS) in solutions with various salts. 
Curiously, even though the adsorption of \ce{I-} increases the CMC as compared to salt \ce{KCl}, the CMC of a solution containing \ce{KI} still remains 
lower than the CMC of a salt free solution.  This shows the dichotomy of electrostatic 
screening and anion adsorption in solutions containing chaotropic anions. 
\begin{table}[h]
\begin{tabular}{c|c}
      \hline
      \hline
                        &  CMCs(mM) \\
      \hline
      no salt           &  59.2  \\
      \ce{KI}           &  52.1  \\
      \ce{KNO3}         &  48.04 \\
      \ce{KBr}          &  46.92 \\
      \ce{KCl}          &  39.94 \\      
      \hline
      \hline      
\end{tabular}
\caption{Summary of the calculated CMCs for SDeS anionic surfactant at $50$~mM salt concentration. 
The $\beta\epsilon$ and $\nu$ parameters are the same as for DMIM.}
\label{tab2}
\end{table}

To account for large hydration radii of small kosmotropic ions such as \ce{Na+} and \ce{F-}, we can slightly increase the size of the unit cell of the lattice model, from $4$~\AA$\,$ to $5$~\AA.  The value of $\epsilon$ then must also be recalculated.  In the case of DMIM surfactant this leads $\epsilon=0.552\, k_B T$.  If we replace $50$~mM \ce{KCl} by $50$~mM \ce{NaF}, we calculate that the CMC of DMIM will increase from $39.94$~mM to $46$~mM. The larger ionic radius hinders the electrostatic screening, resulting in higher CMC in the presence of strongly hydrated kosmotropic ions.

Finally we consider an anionic surfactant sodium dodecyl sulphate~(SDS).   
We find that $\epsilon=0.76\, k_B T$ results in a CMC of $7.76$~mM, very close to the experimental 
value of  $7.8$~mM for a salt free solution of SDS~\cite{DuJa02}. When $15~$mM of \ce{NaCl} is added to solution, we calculate that the CMC for SDS will drop down to $4.35~$mM, very close to the experimental~\cite{DuJa02} value, $4.2~$mM. Unfortunately there is no experimental data for SDS with chaotropic salts to compare with our simulation predictions.

\section{Conclusions}

We have investigated the effect of various salts on the CMCs of ionic surfactants. The Monte Carlo simulations 
of a minimal lattice model were employed to quantitatively predict the CMCs of various ionic surfactants
in different electrolyte solutions. The specific interactions between the 
hydrophobic tails of surfactants and ions were explored based on the insights gained from the earlier theoretical studies of the interfacial tensions~\cite{LeDo09,DoDi10,DoLe10,DoLe12,DoLe13} and the CCCs of hydrophobic colloidal suspensions~\cite{DoLe11}. We find that the kosmotropic anions do not interact with the ionic surfactants, except through steric repulsion and the Coulomb force, while the chaotropic anions interact with the surfactant alkyl tails by the dispersion potential proportional to the ionic polarizability. This is also consistent with the recent experiments which show that strongly hydrated ions are repelled from the hydrophobic groups, while iodide ions are observed next to them~\cite{RaBe13}. The results of simulations are in a good agreement with the available experimental data for cationic surfactants. Using the same model, we are also able to predict the CMCs for anionic surfactants. Unfortunately, at the moment, there is only a very limited experimental data available for these systems 
to compare with our predictions.

\section{Acknowledgments}
This work was partially supported by the CNPq, FAPERGS, FAPESC, INCT-FCx, and by the US-AFOSR under the grant 
FA9550-12-1-0438.

\bibliography{ref.bib}



\end{document}